\documentclass[12pt]{aastex}
\usepackage{apjfonts}

\slugcomment{Accepted for publication in ApJS Spitzer Special Issue 2004}

\begin{document}

\title{Fire and Ice: IRS Mid-IR Spectroscopy of IRAS\,F00183--7111}

\author{H.W.W. Spoon\altaffilmark{1}}
\email{spoon@astro.cornell.edu}
\author{L. Armus\altaffilmark{2}}
\author{J. Cami\altaffilmark{3}}
\author{A.G.G.M. Tielens\altaffilmark{4}}
\author{J.E. Chiar\altaffilmark{5,6}}
\author{E. Peeters\altaffilmark{3}}
\author{J.V. Keane\altaffilmark{5}}
\author{V. Charmandaris\altaffilmark{1}}
\author{P.N. Appleton\altaffilmark{2}}
\author{H.I. Teplitz\altaffilmark{2}}
\author{M.J. Burgdorf\altaffilmark{2}}

\altaffiltext{1}{Cornell University, Astronomy Department, Ithaca, NY 14853}
\altaffiltext{2}{Caltech, Spitzer Science Center, MS 220-6, Pasadena, CA 91125}
\altaffiltext{3}{NASA-Ames Research Center, MS 245-6, Moffett Field, CA 94035}
\altaffiltext{4}{SRON National Institute for Space Research and
Kapteyn Institute, P.O. Box 800, 9700 AV Groningen, The Netherlands}
\altaffiltext{5}{NASA-Ames Research Center, MS 245-3, Moffett Field, CA 94035}
\altaffiltext{6}{SETI Institute, 2035 Landings Drive, Mountain View, CA 94043}

\begin{abstract}
We report the detection of strong absorption and weak emission features 
in the 4--27\,$\mu$m {\it Spitzer-IRS} spectrum of the distant ultraluminous 
infrared galaxy (ULIRG) IRAS F\,00183--7111 (z=0.327). 
The absorption features of CO$_2$ and CO gas, water ice, hydrocarbons
and silicates are indicative of a strongly obscured
(A$_{9.6}$\,$\geq$5.4; A$_{\rm V}$$\geq$90) and complex line of sight 
through both hot diffuse ISM and shielded cold molecular clouds 
towards the nuclear power source. 
From the profile of the 4.67\,$\mu$m CO fundamental vibration mode
we deduce that the absorbing gas is dense (n\,$\sim$10$^6$\,cm$^{-3}$)
and warm (720\,K) and has a CO column density of 
$\sim$10$^{19.5}$\,cm$^{-2}$, equivalent to
N$_{\rm H}$\,$\sim$10$^{23.5}$\,cm$^{-2}$. The high temperature and
density, as well as the small infered size ($<$\,0.03\,pc), locates 
this absorbing gas close to the power source of this region.
Weak emission features of molecular hydrogen, PAHs and Ne$^+$, likely
associated with star formation, are detected against the 9.7\,$\mu$m 
silicate feature, indicating an origin away from the absorbing region. 
Based on the 11.2\,$\mu$m PAH flux, we estimate the star formation 
component to be responsible for up to 30\% of the IR luminosity of the 
system.
While our mid-infrared spectrum shows no tell-tale signs of AGN
activity, the similarities to the mid-infrared spectra of deeply obscured 
sources (e.g. NGC\,4418) and AGN hot dust (e.g. NGC\,1068), as well as
evidence from other wavelength regions, suggest that the power source 
hiding behind the optically thick dust screen may well be a buried AGN.
\end{abstract}


%%%%%%%%%%%%%%%%%%%%%%%%%%%%%%%%%%%%%%%%%%%%%%%%%%%%%%%%%%%%%%%%%%%%%%%%%%
\section{Introduction}
%%%%%%%%%%%%%%%%%%%%%%%%%%%%%%%%%%%%%%%%%%%%%%%%%%%%%%%%%%%%%%%%%%%%%%%%%%

One of the most important discoveries by IRAS was the detection of a
class of galaxies with infrared (8--1000\,$\mu$m) luminosities in
excess of 10$^{12}$\,L$_{\odot}$ and infrared-to-blue ratios 
(L$_{\rm IR}$/L$_{\rm B}$) even higher than for lower luminosity 
infared-bright galaxies. Numerous studies have since established
that ULIRGs are predominantly found in interacting systems and that
their huge luminosities are the result of vigorous, merger-induced 
star formation and/or AGN activity. Spectacular as they may be,
ULIRGs are relatively rare in the Local Universe \citep{Soifer87}.
At higher redshifts though, ULIRGs play an important role in the
measurable star formation density and may account for most of the
far-infrared (FIR) background \citep{Blain02}. As such, the study 
of local members of the ULIRG community may be instrumental in
understanding their high redshift counterparts.

IRAS\,F00183--7111 is a distant ($z$=0.327; D$_L$=1700\,Mpc, assuming 
H$_0$=71\,km\,s$^{-1}$\,Mpc$^{-1}$, $\Omega_M$=0.27,
$\Omega_{\Lambda}$=0.73, $\Omega_K$=0), luminous 
(L$_{\rm IR}$=7$\pm$1$\times$10$^{12}$\,L$_{\odot}$)
ULIRG, discovered by IRAS.
The galaxy has been optically classified as LINER/Seyfert$\geq$1.5
by \citet{Armus89}. Its radio luminosity is in the range of powerful
radio galaxies, far higher than would be expected for starburst
galaxies on the basis of the FIR luminosity 
\citep[q=1.14;][]{Roy97,Norris88}. 
K-band imaging shows the galaxy to have a disturbed morphology
\citep{Rigopoulou99}. 
ISO mid-infrared (MIR) spectroscopy revealed its 5--12\,$\mu$m spectrum
to be markedly different from that of template starbursts and AGNs,
showing instead more similarities to deeply obscured Galactic sources 
\citep{Tran01}. 
Based on the absorbed spectral appearance, \citet{Tran01} conclude
that IRAS\,F00183--7111 is an extremely obscured AGN.

The 5--12\,$\mu$m spectral appearance of IRAS\,F00183--7111 is not 
unique. In a study involving 103 galaxies observed spectroscopically
by ISO \citep{Spoon02}, IRAS\,F00183--7111 is classified along with 
IRAS\,00188--0856, NGC\,4418 and IRAS\,15250+3609 as a class I `ice 
galaxy'; showing ice, silicate and hydrocarbon absorption
features, with little or no sign of PAH emission bands. Based on
this classification, \citet{Spoon02} conclude that the power source
in IRAS\,F00183--7111 may be in the obscured beginnings of star 
formation or AGN activity.

In this Letter we report a new spectrum obtained at a spectral 
resolution of 64--128, extending the observed range from 5--12\,$\mu$m 
to 4--27\,$\mu$m.

%+++++++++++++++++++++++++++++++++++++++++++++++++++++++++++++++++++++
\begin{figure*}[!thb]
\begin{center}
\resizebox{13cm}{!}{\includegraphics{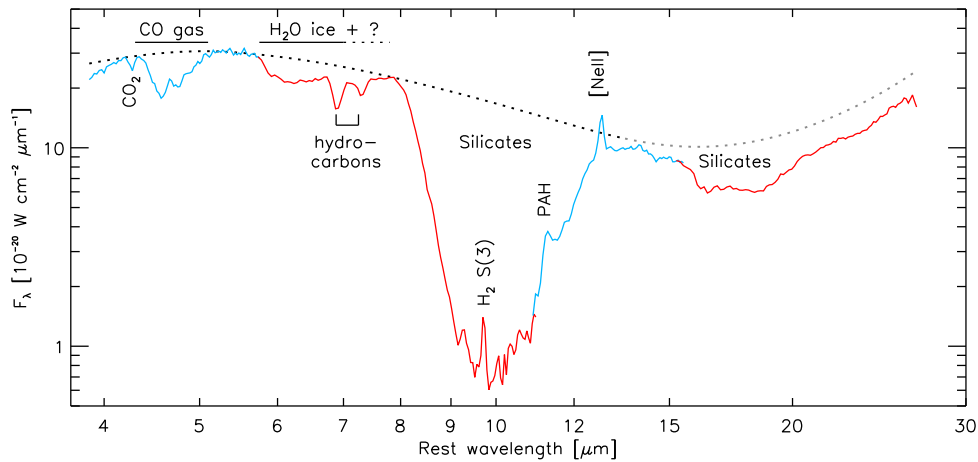}}
\caption{The IRS low resolution spectrum of IRAS\,F00183--7111 
is dominated by strong silicate absorption bands at 9.7 and 
18\,$\mu$m, with weaker absorption bands due to CO$_2$, CO, 
water ice and hydrocarbons visible in the 4.0--7.5\,$\mu$m
range. Weak emission features of H$_2$, PAH and [Ne{\sc ii}] 
are detected at 9.66, 11.2 and 12.8\,$\mu$m, respectively. The 
adopted local continuum is indicated by a {\it dotted line}.
\label{fig1}}
\end{center}
\end{figure*}
%+++++++++++++++++++++++++++++++++++++++++++++++++++++++++++++++++++++

%%%%%%%%%%%%%%%%%%%%%%%%%%%%%%%%%%%%%%%%%%%%%%%%%%%%%%%%%%%%%%%%%%%%%%%%%%
\section{Observations and data reduction}
%%%%%%%%%%%%%%%%%%%%%%%%%%%%%%%%%%%%%%%%%%%%%%%%%%%%%%%%%%%%%%%%%%%%%%%%%%

A low resolution 5--38\,$\mu$m spectrum of IRAS\,F00183--7111 was
obtained on November 14, 2003 using the IRS\footnote{
The IRS was a collaborative venture between Cornell University and 
Ball Aerospace Corporation funded by NASA through the Jet Propulsion 
Laboratory and the Ames Research Center.} spectrometer \citep{Houck04} 
onboard the Spitzer Space Telescope \citep{Werner04}.  The observation
%\dataset[ads/sa.spitzer\#0007556352]{(AOR key 7556352)}
(AOR key 7556352) 
was performed using 60\,sec. ramps in both short-low 
order 2 (SL2; 5.2--7.7\,$\mu$m) and order 1  (SL1; 7.4--14.5\,$\mu$m),
while using 31\,sec. ramps in both long-low order 2 
(LL2; 14.0--21.3\,$\mu$m) 
and order 1 (LL1; 19.5--38.0\,$\mu$m), with a total integration time 
amounting to 15\,minutes.

The spectra were reduced using the IRS pipeline (V\,9.1) at the 
Spitzer Science Center. 
Background emission in our aperture was removed from the SL and LL
images by differencing the two nod positions.
The spectrum of both the target and the calibration star (HD\,105 for
SL and HR\,6348 for LL) were extracted with an aperture which scales
with wavelength. The target spectrum was flux calibrated by dividing
it by the stellar spectrum, followed by multiplication by the adopted 
Cohen stellar model \citep{Cohen03}.
Finally, the long-low section was defringed using the `IRS defringer'. 
For a good match to both LL1 and SL1, the LL2 
spectrum was divided by 1.1. We estimate the absolute flux calibration 
of the resulting spectrum to have a tolerance of less than 20\%.

%%%%%%%%%%%%%%%%%%%%%%%%%%%%%%%%%%%%%%%%%%%%%%%%%%%%%%%%%%%%%%%%%%%%%%%%%%
\section{Analysis}
%%%%%%%%%%%%%%%%%%%%%%%%%%%%%%%%%%%%%%%%%%%%%%%%%%%%%%%%%%%%%%%%%%%%%%%%%%
The IRS spectrum of IRAS\,F00183--7111 (Fig.\,\ref{fig1}) is
dominated by broad and partially overlapping absorption features, 
spanning the entire observable range. Emission features have been
detected only in the 9.5--13\,$\mu$m range.

\subsection{Continuum placement}
Optical depth profiles and the derived physical properties generally 
depend strongly on the assumed local continuum. Here we will adopt
the spline-interpolated continuum as shown in Fig.\,\ref{fig1}.
Although a small contribution of 7.7\,$\mu$m PAH emission to the
7.7\,$\mu$m flux is to be expected, the non-detection of the 
6.2\,$\mu$m PAH band (Sect.\,3.3) indicates that this contribution 
is small ($<$5\%).  We therefore do not correct the 7.7\,$\mu$m 
local continuum for this emission.
At 13.5\,$\mu$m we choose to run the local continuum a factor
exp[$\tau_{\rm sil}$(13.5\,$\mu$m)] above the 13.5\,$\mu$m
flux to correct for the non-zero optical depth in the overlap 
region between the 9.7 and 18\,$\mu$m silicate absorption bands
(see Sect.\,3.2). Beyond 13.5\,$\mu$m, the local continuum is purely 
hypothetical, designed only to give a reasonable fit to the 
18\,$\mu$m silicate feature.

%+++++++++++++++++++++++++++++++++++++++++++++++++++++++++++++++++++++
\begin{figure*}
\begin{center}
\resizebox{13.7cm}{!}{\includegraphics{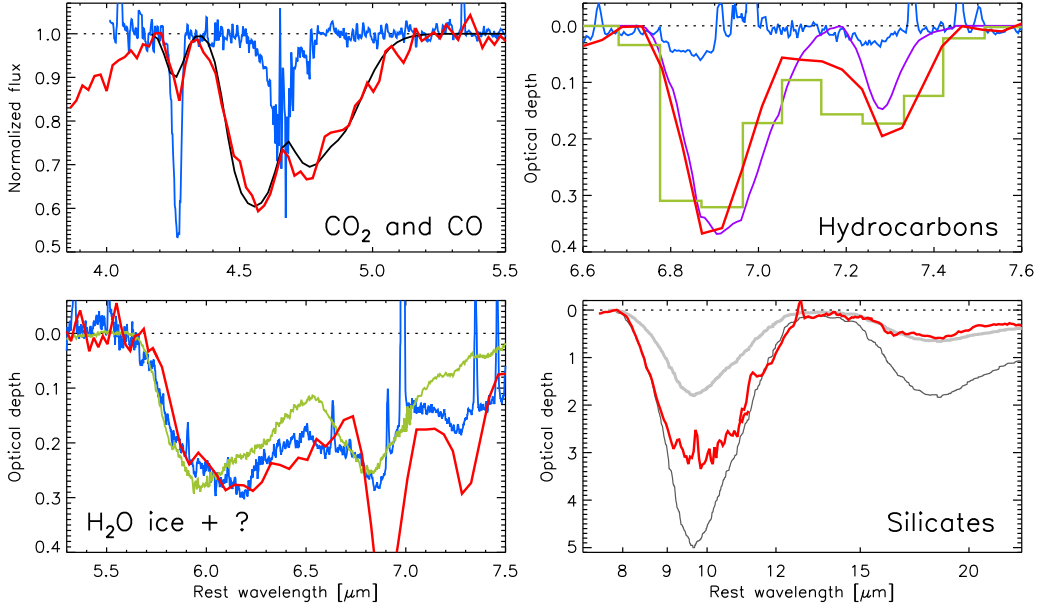}}
\caption{Absorption features in the spectrum of IRAS\,F00183--7111 
({\it red}). 
{\it Upper left panel:}
comparison of the CO$_2$ and CO normalized flux 
profiles of IRAS\,F00183--7111 and Sgr\,A$^*$ ({\it blue}). 
The best gas model fit to the IRAS\,F00183--7111 profile is shown 
in {\it black}. Our data cannot exclude the CO$_2$ feature to be
(partially) due to solid CO$_2$ instead.
{\it Lower left panel:} 
comparison of the 5.5--7.5\,$\mu$m optical depth profile of 
IRAS\,F00183--7111 to that of Sgr\,A$^*$ ({\it blue}; 
multiplied by 1.4) and W\,33A ({\it green}; divided by 6.2).
{\it Top right panel:} 
comparison of the 6.6--7.6\,$\mu$m hydrocarbon optical depth 
profiles of IRAS\,F00183--7111,  Sgr\,A$^*$ ({\it blue}) and
NGC\,4418 ({\it green histogram}; divided by 3). A laboratory 
profile of HAC \citep{Furton99} is shown in {\it purple}.
{\it Bottom right panel:} 
silicate optical depth profile of IRAS\,F00183--7111. The 
{\it grey lines} show two fits of the baseline-subtracted silicate 
profile of GCS\,3I to the observed IRAS\,F00183--7111 
profile. The {\it dark grey line} represents the best fit to 
the wings of the 9.7\,$\mu$m feature and the {\it light grey line} 
the best fit to the 18\,$\mu$m feature.
\label{fig2}}
\end{center}
\end{figure*}
%+++++++++++++++++++++++++++++++++++++++++++++++++++++++++++++++++++++

\subsection{Absorption features}
Table\,\ref{tab1} gives an overview of the absorption features
detected in the spectrum of IRAS\,F00183--7111; the associated 
optical depth profiles are presented in Fig.\,\ref{fig2}.
In our analysis of the absorption features, we have used the 
spectra of extinguished Galactic lines of sight 
(the Galactic center sources GCS\,3I and 
Sgr\,A$^*$ and the embedded massive protostar W\,33A), as well 
as one extragalactic line of sight (the nucleus of NGC\,4418) 
for comparison.

The broad double branched feature centered at 4.67\,$\mu$m 
is identified with the P and R branches of the fundamental
vibration mode of CO gas. Using the isothermal plane parallel 
LTE gas models of \citet{Cami02}, a best fit to the observed 
profile is found for a model with a gas temperature of 720\,K,
an intrinsic linewidth of 50\,km/s and a column density of
10$^{19.5}$\,cm$^{-2}$.
The derived temperature is far higher than found towards 
Galactic ISM lines of sight, consistent with the clearly
larger width of the feature. A comparison of the model to 
the data is shown in Fig.\,\ref{fig2}. For Galactic sources,
a narrow absorption band at 4.26\,$\mu$m is normally 
attributed to CO$_2$ ice. However, despite extensive 
comparison to Galactic ice sources, neither the peak position 
(4.273\,$\mu$m) nor the red wing of the feature in 
IRAS\,F00183--7111 is found to be consistent with solid state 
CO$_2$. Using our gas models, a reasonable fit is obtained for 
gas phase CO$_2$ at a temperature $\leq$110\,K.

Another broad absorption complex extends from 5.7 to almost 
7.8\,$\mu$m, with peaks at 6.15, 6.85 and 7.25\,$\mu$m.
The peak positions and profiles appear very similar
to those found for Sgr\,A$^*$ \citep{Chiar00}. The blue wing
(5.7--5.9\,$\mu$m) also matches the profile observed towards
W\,33A \citep{Gibb00}. Beyond 6.0\,$\mu$m, the profiles of
IRAS\,F00183--7111 and Sgr\,A$^*$ are markedly different from 
that of W\,33A, peaking at 6.15--6.20\,$\mu$m instead of 
6.0\,$\mu$m. While the W\,33A profile is consistent with 
the profile of water ice, the shift in peak position for 
IRAS\,F00183--7111 and Sgr\,A$^*$ indicates the presence
of strong additional absorption, likely due to hydrocarbons
\citep{Keane04}.
The two absorption bands at 6.85 and 7.25\,$\mu$m are 
identified with the C--H stretching mode of hydrocarbons 
\citep[e.g.][]{Furton99}. The only
other two sources for which these bands have been detected 
are  Sgr\,A$^*$ \citep{Chiar00} and the nucleus of NGC\,4418 
\citep{Spoon01}. As shown in 
Fig.\,\ref{fig2}, the features compare well with those seen 
in a laboratory spectrum of hydrogenated amorphous carbons 
(HAC) \citep{Furton99}.

Beyond 8\,$\mu$m, the IRAS\,F00183--7111 spectrum shows two 
characteristic broad absorption bands due to silicates. To 
estimate the optical depth in both features, we use a
baseline-subtracted silicate absorption profile of GCS\,3I 
\citep{Chiar04}. Assuming the contribution of starburst-related
emission to be negligible, a best fit to the blue wing of the 
9.7\,$\mu$m feature is obtained for an optical depth of $\sim$5 
(see Fig.\,\ref{fig2}). This fit also gives reasonable result 
for the red wing, but fails to fit the center of the feature.
As will be discussed in Sect.\,3.3, the latter is an indication 
for the presence of a separate, extended, low intensity, 
relatively unobscured emission component, filling in
the bottom of the feature.
The derived optical depth turns into a lower limit if the 
contribution from starburst emission is higher than assumed.
Note that the fit to the 9.7\,$\mu$m feature overpredicts 
the depth of the 18\,$\mu$m band by more than a factor of 3.
This either points to the presence of a strong starburst 
continuum filling in the 18\,$\mu$m silicate feature, or 
to the effects of radiative transfer in warm silicates.
In either this case the true silicate optical depth may
be far higher than inferred from our fit.

%+++++++++++++++++++++++++++++++++++++++++++++++++++++++++++++++++++++
\begin{table}
\caption{Absorption features\label{tab1}}
\begin{tabular}{lccccc}
\tableline
\tableline
Species & $\lambda_{\rm rest}$ &
$\tau_{\rm max}$ & $\tau_{\rm int}$\tablenotemark{a} & $N$ & $N_{\rm H}$ \\
 & [$\mu$m] &  & [cm$^{-1}$] & [10$^{18}$ cm$^{-2}$] & [10$^{22}$ cm$^{-2}$] \\
\tableline
CO$_2$ gas  & 4.26 &           &        & 0.05 &  5\tablenotemark{b}\\
CO     gas  & 4.67 &           &        & 32   &  32\tablenotemark{c}\\
H$_2$O ice + ?  & 6.15 & 0.28  &   75   &      &   \\
`HAC'       & 6.85 & 0.36      &   15   & 11\tablenotemark{d} &   \\
`HAC'       & 7.25 & 0.19      &    6   & 11\tablenotemark{d} &   \\
silicates   & 9.6  & $\geq$5.0 & $\geq$1130&& $\geq$17\tablenotemark{e}\\
silicates   & 18   & 0.57\tablenotemark{f}&129\tablenotemark{f} && \\
\tableline
\end{tabular}
\tablenotetext{a}{$\tau_{\rm int}$=$\int\tau_{\rm \nu}$d$\nu$}
\tablenotetext{b}{Assuming N$_H$$\sim$10$^6$ N$_{\rm CO_2}$}
\tablenotetext{c}{Assuming all carbon to be locked up in CO, 
N$_H$$\sim$10$^4$ N$_{\rm CO}$}
\tablenotetext{d}{Computed using A(6.85\,$\mu$m)=1.35$\times$10$^{-18}$ cm/mol.
and A(7.25\,$\mu$m)=0.6$\times$10$^{-18}$ cm/mol.; \citet[]{Wexler67}}
\tablenotetext{e}{Assuming A$_{9.6}$/A$_{\rm V}$=0.06 and
N$_{\rm H}$=1.9$\times$10$^{21}$\,A$_{\rm V}$\,cm$^{-2}$}
\tablenotetext{f}{Assuming our choice for the local 14--27\,$\mu$m continuum}
\end{table}
%+++++++++++++++++++++++++++++++++++++++++++++++++++++++++++++++++++++

\subsection{Emission features}

The spectrum of IRAS\,F00183--7111 (Fig.\,\ref{fig1})
reveals the presence of several emission features: 
9.66\,$\mu$m H$_2$ 0--0 S(3), 11.2\,$\mu$m PAH and the 
12.7--12.8\,$\mu$m blend of PAH and [Ne{\sc ii}]. 
Interestingly, their presence in the center and red flank of the 
optically thick 9.7\,$\mu$m silicate band suggests a common origin 
away from the absorbing medium. Also the weak continuum,
filling in the silicate feature, may be associated with these
emission features. We list the measured line strengths as well as
upper limits for lines commonly detected in star formation 
environments in Table\,\ref{tab2}.

Given the clear detection of the 11.2\,$\mu$m PAH band, the 
non-detection of the 6.2\,$\mu$m PAH band may seem surprising.
However, as \citet{Hony01} have shown for Galactic ISM lines
of sight, the ratio 6.2PAH/11.2PAH is variable and ranges from
1 to 3 for H{\sc ii} regions. At a value of $<$0.83,
the ratio for IRAS\,F00183--7111 is outside the range for 
Galactic H{\sc ii} regions,
but still above the values found for the galaxies NGC\,5195 
\citep[0.57;][]{Boulade96} and NGC\,3226 \citep[0.35;][]{Appleton04}.
Interestingly, in another study of  H{\sc ii} regions,
\citet{Vermeij02} find that sources with 6.2PAH/11.2PAH$<$1
all happen to be low metallicity objects (e.g. SMC\,B1\#1 and 
positions in 30\,Dor).

A rough estimate of the IR flux associated with the star formation 
component can
be made using conversion factors L(11.2PAH)/L(IR)=0.0014 and
L(6.2PAH)/L(IR)=0.0034, as proposed by \citet{Soifer02} for M\,82
and \citet{Peeters04} for normal and starburst galaxies, respectively. 
The resulting contribution from the unobscured star formation 
component to the IR luminosity budget of IRAS\,F00183--7111 hence 
may range from $\leq$10\% (based on the 6.2\,$\mu$m PAH upper limit) 
to 30\% (based on the 11.2\,$\mu$m PAH flux). 

Finally, we investigated the nature of the [Ne{\sc ii}] flux by
comparing it to the measured 11.2\,$\mu$m PAH flux and the 
[Ne{\sc ii}]/11.2PAH ratio for a sample of (compact) H{\sc ii} 
regions \citep{Peeters02}. We find that the observed ratio for 
IRAS\,00183--7111 (0.6) compares well to the Galactic values. The
observed [Ne{\sc ii}] flux may hence be ascribed completely to 
unobscured star formation.

%+++++++++++++++++++++++++++++++++++++++++++++++++++++++++++++++++++++
\begin{table}
\caption{Emission features\label{tab2}}
\begin{tabular}{lcc|lcc}
\tableline\tableline
Species & $\lambda_{\rm rest}$  & flux & 
Species & $\lambda_{\rm rest}$  & flux \\
        & [$\mu$m]              & [10$^{-21}$ W cm$^{-2}$] &
        & [$\mu$m]              & [10$^{-21}$ W cm$^{-2}$] \\
\tableline
PAH               & 6.22  & $<$3    & PAH              & 12.7  & 2\tablenotemark{a}\\
$[$Ar{\sc ii}$]$  & 6.99  & $<$1.3  &$[$Ne{\sc ii}$]$  & 12.8  & 6\tablenotemark{a} \\
H$_2$ 0--0 S(3)   & 9.66  & 0.6     &$[$Ne{\sc v}$]$   & 14.3  & $<$1.5  \\
PAH               & 11.2  & 3.6     &$[$Ne{\sc iii}$]$ & 15.6  & $<$3    \\
\tableline
\end{tabular}
\tablenotetext{a}{The [Ne{\sc II}] flux was measured by fitting a 
gauss profile to the peak and red wing of the 12.55--12.95\,$\mu$m 
emission complex. The remaining flux in the blue wing (25\%) is 
ascribed to the 12.7\,$\mu$m PAH feature.}
\end{table}
%+++++++++++++++++++++++++++++++++++++++++++++++++++++++++++++++++++++

%%%%%%%%%%%%%%%%%%%%%%%%%%%%%%%%%%%%%%%%%%%%%%%%%%%%%%%%%%%%%%%%%%%%%%%%%%
\section{Discussion and conclusions}
%%%%%%%%%%%%%%%%%%%%%%%%%%%%%%%%%%%%%%%%%%%%%%%%%%%%%%%%%%%%%%%%%%%%%%%%%%

We have compared the wealth of emission and absorption features in the 
exotic MIR spectrum of IRAS\,F00183--7111 to (extra)galactic lines of
sight displaying similar features. 
The picture that emerges is that of an energetically dominant nuclear
power source, buried deeply behind two obscuring shells. The inner
shell consists of warm gas 
(T$\sim$720\,K; N$_{\rm H}$\,=10$^{23.5}$\,cm$^{-2}$; Table\,\ref{tab1}), 
giving rise to absorption bands of warm CO gas and hydrocarbons against 
an otherwise featureless hot MIR continuum. This absorbing gas has to
be dense (n\,$>$3$\times$10$^6$\,cm$^{-3}$) as well in order to give
rise to high-J level absorption, which limits the size of the
absorbing region to $<$\,0.03\,pc. The high temperature and density as
well as the small scale size locates this absorbing gas to close to
the power source of this region. The outer shell is far colder 
and is responsible for the absorption features of CO$_2$
ice/gas, water ice and silicates. Given the depth of the
9.7\,$\mu$m silicate band ($\tau_{9.7}$$\geq$5), the column density 
of the outer shell amounts to at least N$_{\rm H}$\,=10$^{23}$\,cm$^{-2}$
(Table\,\ref{tab1}). 
The emission lines from molecular hydrogen, PAHs and Ne$^+$, as well 
as the weak continuum filling in the 9.7\,$\mu$m silicate feature, 
likely originate outside this shell structure. Based on the strength of 
the 11.2\,$\mu$m PAH feature, this star formation component may be
responsible for up to 30\% of the IR luminosity of the system.

\begin{figure}
\resizebox{\hsize}{!}{\includegraphics{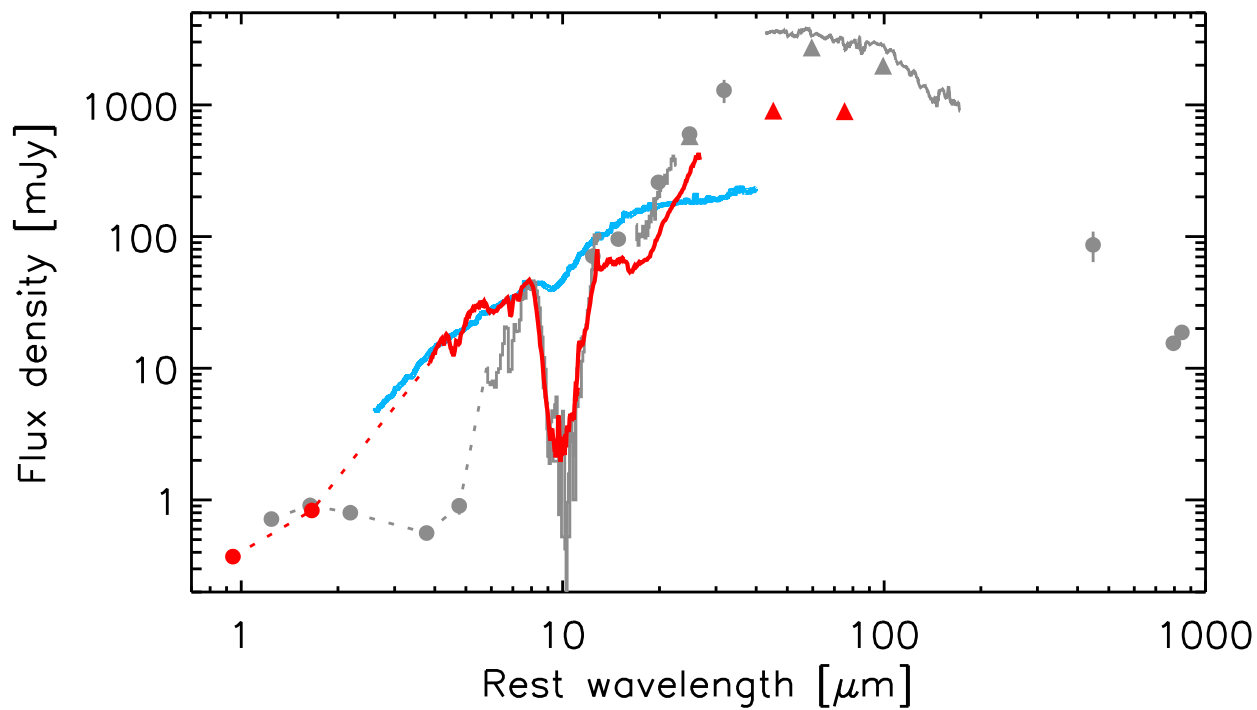}}
\caption{Comparison of the IR SEDs of IRAS\,F00183--7111
({\it red}), NGC\,4418 \citep[{\it grey};][]{Spoon01} and the nucleus of 
NGC\,1068 \citep[{\it blue}; smoothed;][]{Sturm00}. The latter two spectra 
have been scaled to match the 8\,$\mu$m flux of IRAS\,F00183--7111.
IRAS FSC fluxes are denoted by {\it triangles}. The near-IR data for 
IRAS\,F00183--7111 have been published by \citet{Rigopoulou99}.
\label{fig3}}
\end{figure}

Optical and radio observations indicate that the power source
hiding in the nucleus is most likely an AGN. The optical evidence
comes from an emission line study of \citet{Armus89}, who classify
IRAS\,F00183--7111 as a LINER/Seyfert$\geq$1.5. Convincing evidence 
is also presented by \citet{Roy97}, who show that the radio power 
of IRAS\,F00183--7111 is far higher than expected on the basis of 
the radio-FIR correleation for starburst galaxies. Radio VLBI 
observations further indicate the nucleus to be unresolved on the 
0.025$\arcsec$ (200\,pc) scale (R.P.\,Norris, priv. comm.).
Our IRS data is in agreement with these findings. In a diagram 
of  L(6.2PAH)/L(FIR) versus L(6.2cont)/L(FIR), the so-called 
`MIR-FIR diagram' \citep{Peeters04}, IRAS\,F00183--7111 is found at 
a position intermediate between a deeply obscured source 
(e.g. NGC\,4418) and AGNs with a clear line of sight to their 
AGN-heated hot dust (e.g. NGC\,1068). 
The latter can also be infered from Fig.\,\ref{fig3}, where the 
IR spectrum of IRAS\,F00183--7111 shows signatures of both
an AGN-like hot dust continuum \citep[NGC\,1068;][]{Sturm00} 
and a deeply obscured source \citep[NGC\,4418;][]{Spoon01}

In a follow-up study, high-resolution IRS spectroscopy will be 
used to search for direct mid-infrared AGN tracers such as 
14.3\,$\mu$m and 24.3\,$\mu$m [Ne{\sc v}].

%+++++++++++++++++++++++++++++++++++++++++++++++++++++++++++++++++++++
%+++++++++++++++++++++++++++++++++++++++++++++++++++++++++++++++++++++

\acknowledgments
The authors would like to thank Lou Allamandola and the referee, 
Pascale Ehrenfreund, for helpful comments.
Support for this work was provided by NASA through the Spitzer Space
Telescope Fellowship Program, through Contract Number 1257184 
issued by the Jet Propulsion Laboratory, California Institute of
Technology under NASA contract 1407.

\end{document}